\pdfoutput=1
\documentclass{article}
\usepackage{spconf,amsmath,graphicx}

\title{Comparing Baseline and Day-1 Diffusion MRI Using Multimodal Deep Embeddings for Stroke Outcome Prediction}

\name{Sina Raeisadigh$^{1}$, Myles Joshua Toledo Tan$^{2}$, Henning Müller$^{3}$, Abderrahmane Hedjoudje$^{4}$}

\address{$^{1}$ Department of Computer Science, University of Geneva, Switzerland \\
         $^{2}$ Department of Electrical \& Computer Engineering, University of Florida, FL, USA \\
         $^{3}$ Service of Medical Informatics, University Hospital of Geneva, Switzerland \\
         $^{4}$ Department of Imaging and Medical Informatics, University of Geneva, Switzerland}
\begin{document}

\maketitle
\begin{abstract}
This study compares baseline (J0) and 24-hour (J1) diffusion magnetic resonance imaging (MRI) for predicting three-month functional outcomes after acute ischemic stroke (AIS). Seventy-four AIS patients with paired apparent diffusion coefficient (ADC) scans and clinical data were analyzed. Three-dimensional ResNet-50 embeddings were fused with structured clinical variables, reduced via principal component analysis (\(\leq\)12 components), and classified using linear support vector machines with eight-fold stratified group cross-validation. J1 multimodal models achieved the highest predictive performance (AUC = 0.923 ± 0.085), outperforming J0-based configurations (AUC \(\leq\) 0.86). Incorporating lesion-volume features further improved model stability and interpretability. These findings demonstrate that early post-treatment diffusion MRI provides superior prognostic value to pre-treatment imaging and that combining MRI, clinical, and lesion-volume features produces a robust and interpretable framework for predicting three-month functional outcomes in AIS patients.
\end{abstract}

\section{Introduction}
Acute ischemic stroke (AIS) is a leading cause of death and long-term disability worldwide~\cite{feigin2021global}. Reliable early prediction of post-stroke functional outcome is critical for treatment planning, rehabilitation, and patient selection in clinical trials~\cite{heiss2017neuroimaging}. The three-month modified Rankin Scale (mRS) remains the standard measure of disability~\cite{banks2007stroke}, yet accurate early prediction remains challenging due to heterogeneity in stroke mechanisms, treatment response, and recovery. Such a personalized prognostic model can support treatment decisions, rehabilitation planning, and adaptive trial enrollment, addressing a major unmet clinical need in stroke management.

Diffusion-weighted MRI (DWI) and its quantitative map, the apparent diffusion coefficient (ADC), are sensitive to early ischemic injury~\cite{warach1992fast}. Baseline (J0) ADC reflects the infarct core, while day-1 (J1) imaging captures infarct evolution, re-perfusion, and secondary injury processes~\cite{campbell2021acute, goyal2020challenging}. Although both time points are  acquired routinely, their relative prognostic value for long-term functional outcome remains unclear.

Traditional models combining clinical scores such as age, NIHSS (National Institutes of Health Stroke Score), and pre-stroke mRS achieve moderate predictive accuracy (AUC $\approx$ 0.75–0.80)~\cite{fiebach2002ct}, but underuse imaging biomarkers reflecting dynamic tissue changes. Deep learning enables automated extraction of high-dimensional image representations~\cite{esteva2019guide, nielsen2018prediction}, and recent multimodal models integrating imaging and clinical data have improved prediction~\cite{kim2023deep, ma2024transformer}. However, most prior studies focus solely on baseline imaging.

To our knowledge, this is the first study to directly compare baseline (J0) and early follow-up (J1) diffusion MRI for predicting three-month functional outcome in AIS using a unified multimodal deep-embedding framework. We hypothesize that post-treatment (J1) imaging, combined with clinical and lesion-volume features, can improve predictive performance and interpretability. Our framework integrates 3D ResNet-derived ADC embeddings with structured clinical variables and applies Principal Component Analysis (PCA) for dimensionality reduction and linear SVM classification to enable transparent, data-efficient outcome prediction.

Functional outcome prediction after AIS has relied on clinical scoring and regression models. Baseline neurological severity (NIHSS), age, and pre-stroke mRS remain the strongest predictors of three-month outcomes~\cite{heiss2017neuroimaging}. Logistic or Cox models using these variables achieve moderate accuracy (AUC $\approx$ 0.75–0.80)~\cite{fiebach2002ct, wouters2018prediction}, but lack sensitivity to imaging biomarkers reflecting infarct evolution or treatment response.

Machine learning and deep learning methods can further improve prediction by combining imaging and clinical features~\cite{lecun2015deep,kim2023deep, esteva2019guide}. CNNs and 3D architectures enable automated feature extraction from ADC maps~\cite{winzeck2018isles, liu2023functional}, and multimodal fusion models show superior accuracy~\cite{ma2024transformer}. Yet most prior work evaluates only baseline scans.

To mitigate redundancy in deep embeddings, PCA often offers efficient dimensionality reduction and improved generalization~\cite{wold1987principal}. Our study builds on this by comparing J0 and J1 imaging, integrating deep volumetric and clinical features, and employing PCA-based fusion with linear SVM classification for interpretable AIS outcome prediction.
\section{Methods}
\subsection{Study Cohort and Clinical Data}
\begin{figure}[t]
    \centering
    \includegraphics[width=1.00\linewidth]{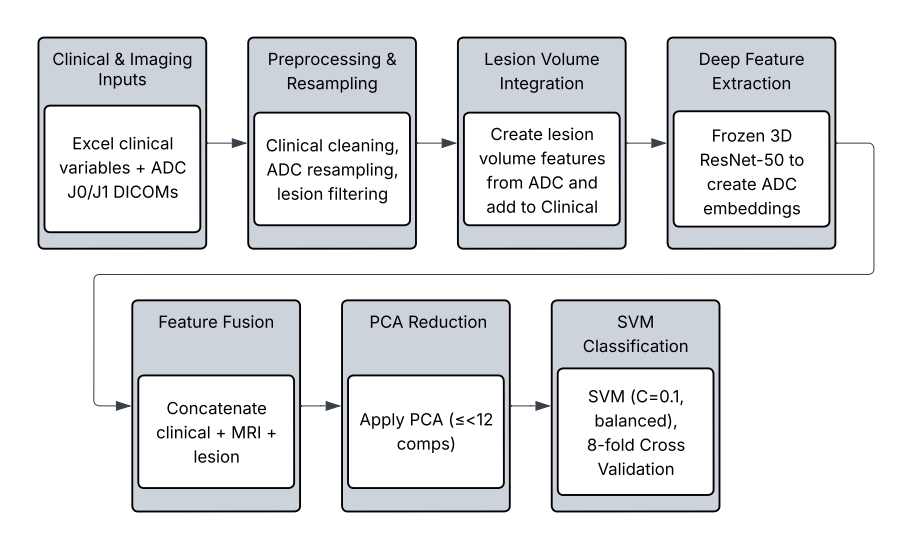}
    \caption{
    Overview of the proposed multimodal pipeline combining clinical data and ADC imaging (J0, J1). 
    Lesion volumes and deep MRI embeddings are fused, reduced via PCA, and classified using a linear SVM.
    }
    \label{fig:pipeline}
\end{figure}

74 patients with AIS who underwent DWI MRI at J0 and J1 were retrospectively selected from a prospective registry under institutional approval. Inclusion required paired ADC volumes, complete clinical records (NIHSS subscores, pre-stroke mRS), and available three-month mRS. Demographics, vascular risk factors, pre-mRS, and NIHSS (J0, J1) were included. Both anterior and posterior strokes were included; outcomes were 55.4\% favorable (mRS $\leq$ 1) and 44.6\% unfavorable (mRS $>$ 1). Missing values were imputed using median substitution, and all data were anonymized.
The overall data processing and modeling workflow is illustrated in Figure~\ref{fig:pipeline}, which summarizes each step from image preprocessing and lesion extraction to feature fusion and classification.
\subsection{MRI Processing and Lesion Features}
ADC maps reconstructed from 1.5\,T and 3\,T DWI were resampled via trilinear interpolation to a unified 3D resolution of $24 \times 256 \times 256$ voxels. 

\begin{figure}[t]

    \centering
    \includegraphics[width=0.47\textwidth]{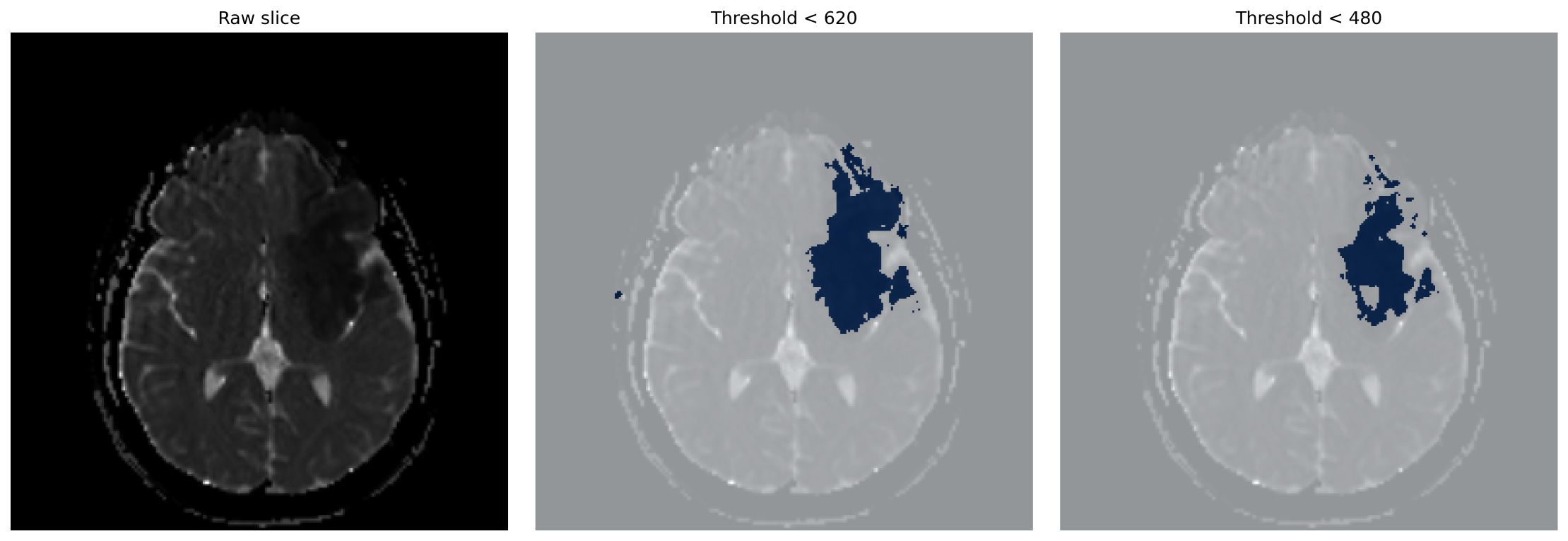}
    \caption{Thresholding of lesion-like regions in ADC maps at different intensity levels. (Left) Raw slice, (Middle) Threshold $<$ 620, (Right) Threshold $<$ 480.}
    \label{fig:thresholding}
\end{figure}

Lesion-like regions were segmented using percentile-based intensity thresholds (480 and 620~$\times 10^{-6}$~mm$^2$/s), followed by morphological filtering and removal of small components ($<$150~voxels). An example of the thresholding process at different intensity levels is shown in Figure~\ref{fig:thresholding}, where (left) shows the raw slice, (middle) shows the result for threshold $<$ 620, and (right) shows the result for threshold $<$ 480. Lesion volume ($V_{\text{lesion}} = N_{\text{voxels}} \times V_{\text{voxel}}$) was log-transformed and concatenated with clinical data.

\subsection{Deep Feature Extraction and Fusion}
Volume features are extracted with a pretrained 3D ResNet-50 (MedicalNet~\cite{chen2019med3d}) implemented in MONAI~\cite{cardoso2022monai} (Medical Open Network for AI). Separate branches processed J0 and J1 ADC input, with global average pooling producing 2048-D embeddings projected to 32–256 units. Networks operated in frozen mode, and embeddings were concatenated with clinical and lesion-volume features. Feature importance is derived from standard SVM coefficients and MRI saliency maps.
\subsection{Dimensionality Reduction and Classification}
Fused multimodal embeddings were reduced using PCA, retaining up to 12 components ($>$95\% variance). PCA was fit on training folds only. Reduced features were classified via linear SVMs ($C{=}0.1$) with class-balanced weighting and Platt scaling~\cite{cortes1995support}. Models include (1) MRI-only, (2) clinical-only, and (3) multimodal (with/without lesion volumes).
\subsection{Evaluation and Implementation}
Performance was evaluated with eight-fold StratifiedGroupKFold cross-validation, reporting mean~$\pm$~SD (Standard Deviation) for AUC, accuracy, and F1-score. All analyses use PyTorch~\cite{paszke2019pytorch}, MONAI~\cite{cardoso2022monai}, and scikit-learn~\cite{pedregosa2011scikit_learn} (Python~3.10) on a Tesla V100 GPU.
Model performance was compared across eight stratified group cross-validation folds. The Wilcoxon signed-rank test~\cite{wilcoxon1945individual} was used to assess paired differences in validation AUC between J0 and J1 multimodal configurations. A one-sided test was applied to test the hypothesis that J1 models outperform J0 models.
\section{Results}

\subsection{Comparison of J0 and J1 MRI Models}
MRI-only models use 3D ResNet-50 embeddings from J0/J1 ADC, with 12 PCA components, classified with class-balanced linear SVMs. In Table~\ref{tab:mri_results}, J1 outperformed J0 across metrics (AUC $0.714 \pm 0.105$ vs.\ $0.540 \pm 0.263$; accuracy $0.633$ vs.\ $0.433$; F1 $0.560$ vs.\ $0.396$), indicating 24~h diffusion changes provide stronger prognostic signal than baseline.
\begin{table}[t]
\centering
\caption{Performance of MRI-only models for predicting three-month mRS ($\leq$1 vs.\ $>$1). 
Values are mean $\pm$ standard deviation across folds.}
\label{tab:mri_results}
\setlength{\tabcolsep}{4.2pt} 
\scriptsize                  
\begin{tabular}{lccc}
\hline
\textbf{Model} & \textbf{AUC} & \textbf{Accuracy} & \textbf{F1-score} \\
\hline
J0 ADC (MRI-only) & $0.540 \pm 0.263$ & $0.433 \pm 0.147$ & $0.396 \pm 0.110$ \\
J1 ADC (MRI-only) & $\mathbf{0.714 \pm 0.105}$ & $\mathbf{0.633 \pm 0.066}$ & $\mathbf{0.560 \pm 0.128}$ \\
\hline
\end{tabular}
\end{table}
\subsection{Clinical and Multimodal Model Performance}
\begin{figure}[t]
\centering
\includegraphics[width=1.00\linewidth]{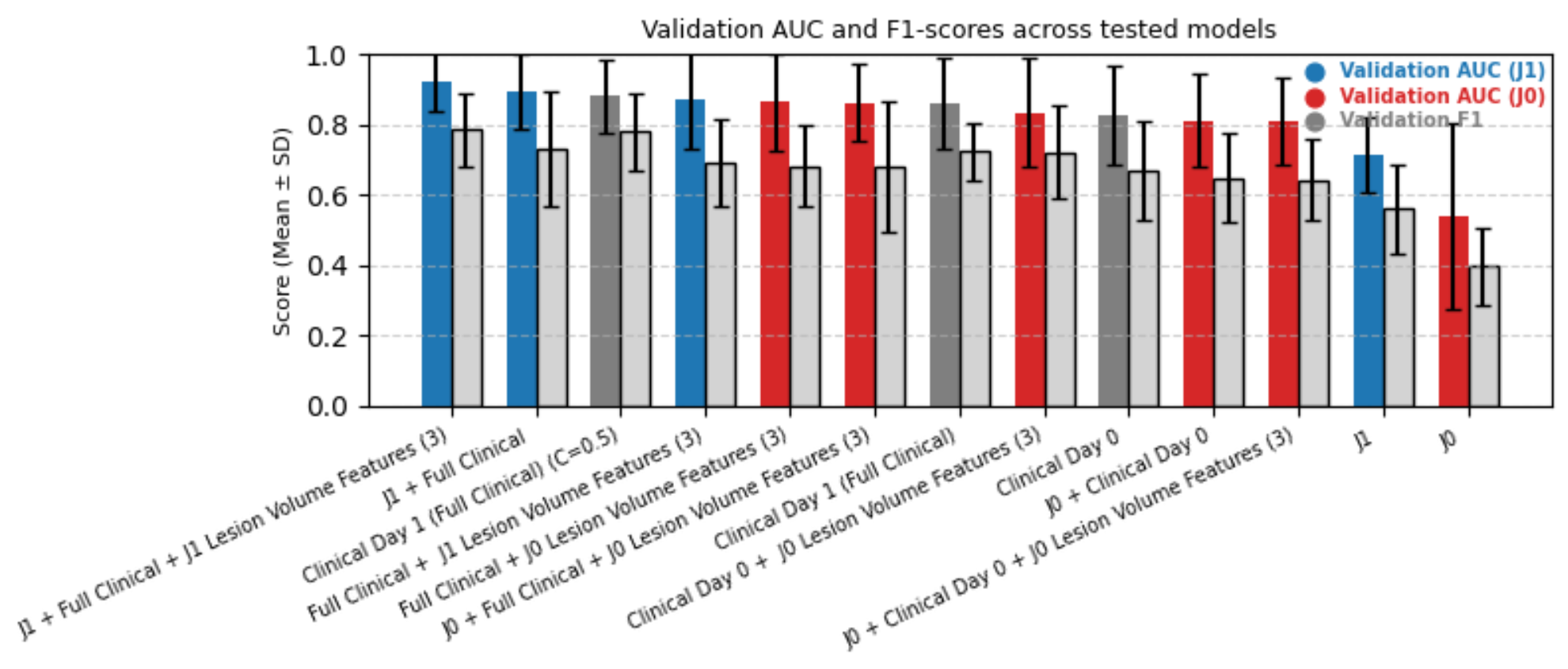}
\caption{Validation AUC and F1 across tested models. J1 multimodal configurations outperformed J0 and clinical baselines.}
\label{fig:perf_summary}
\end{figure}

\begin{table*}[t]
\centering
\caption{Summary of multimodal configurations (mean $\pm$ Standar across 8 folds). PCA dimension fixed at 12 components.}
\label{tab:multimodal_results}

\scriptsize
\setlength{\tabcolsep}{10pt}

\begin{tabular}{p{4cm}cccccc}
\hline
\textbf{Configuration} & \textbf{Val AUC} & \textbf{Val Acc} & \textbf{Val F1} & \textbf{Train AUC} & \textbf{Train Acc} & \textbf{Train F1} \\
\hline
J1 + Full Clinical + J1 Lesion Volume 
& \textbf{0.923 $\pm$ 0.085} 
& \textbf{0.824 $\pm$ 0.083} 
& \textbf{0.786 $\pm$ 0.104} 
& 0.941 $\pm$ 0.011 
& 0.898 $\pm$ 0.018 
& 0.884 $\pm$ 0.021 \\

J1 + Full Clinical 
& 0.894 $\pm$ 0.106 
& 0.794 $\pm$ 0.096 
& 0.731 $\pm$ 0.163 
& 0.918 $\pm$ 0.015 
& 0.888 $\pm$ 0.025 
& 0.876 $\pm$ 0.031 \\

Clinical (Day 1 only)
& 0.882 $\pm$ 0.105 
& 0.808 $\pm$ 0.101 
& 0.780 $\pm$ 0.111 
& 0.912 $\pm$ 0.010 
& 0.853 $\pm$ 0.023 
& 0.832 $\pm$ 0.024 \\

Full Clinical + J0 Lesion Volume
& 0.864 $\pm$ 0.136 
& 0.740 $\pm$ 0.106 
& 0.683 $\pm$ 0.117 
& 0.884 $\pm$ 0.012 
& 0.824 $\pm$ 0.039 
& 0.793 $\pm$ 0.050 \\

J0 + Full Clinical + J0 Lesion Volume
& 0.863 $\pm$ 0.111 
& 0.768 $\pm$ 0.113 
& 0.682 $\pm$ 0.185 
& 0.887 $\pm$ 0.015 
& 0.840 $\pm$ 0.021 
& 0.811 $\pm$ 0.028 \\
\hline
\end{tabular}
\end{table*}
Models trained only on clinical features achieved moderate performance (AUC $= 0.882 \pm 0.105$, accuracy $= 0.808 \pm 0.101$, F1-score $= 0.780 \pm 0.111$), confirming the predictive value of age, pre-stroke mRS, and NIHSS scores. Integrating imaging features further improves results in terms of AUC. When combining J1 MRI embeddings with full clinical vectors, the multimodal model reached AUC $= 0.894 \pm 0.106$, accuracy $= 0.794 \pm 0.096$, and F1-score $= 0.731 \pm 0.163$. Adding lesion-volume metrics derived from ADC thresholding provided additional gains, achieving AUC $= 0.923 \pm 0.085$, accuracy $= 0.824 \pm 0.083$, and F1-score $= 0.786 \pm 0.104$—the highest overall performance among all configurations (Table~\ref{tab:multimodal_results}). The final J1 multimodal configuration (J1 + clinical + J1 lesion volume) achieved higher validation AUCs compared to the corresponding J0 configuration (J0 + clinical + J0 lesion volume; 0.923 ± 0.079 vs.\ 0.811 ± 0.115; Wilcoxon signed-rank test, $p$ = 0.0078), confirming the superior prognostic value of early post-treatment diffusion MRI.
Figure~\ref{fig:perf_summary} shows validation performance across models. J1 multimodal configurations outperformed J0 and clinical-only baselines, with lesion-volume features providing the highest AUC and F1-scores alone.

PCA is critical for stable training with our small dataset. Without PCA, SVMs show unstable convergence and large variance. A 12-component configuration produced the most consistent validation AUCs while preserving interpretability.
\subsection{Feature Importance and Explainability}
\begin{figure}[t]
    \centering
    \includegraphics[width=1.00\linewidth]{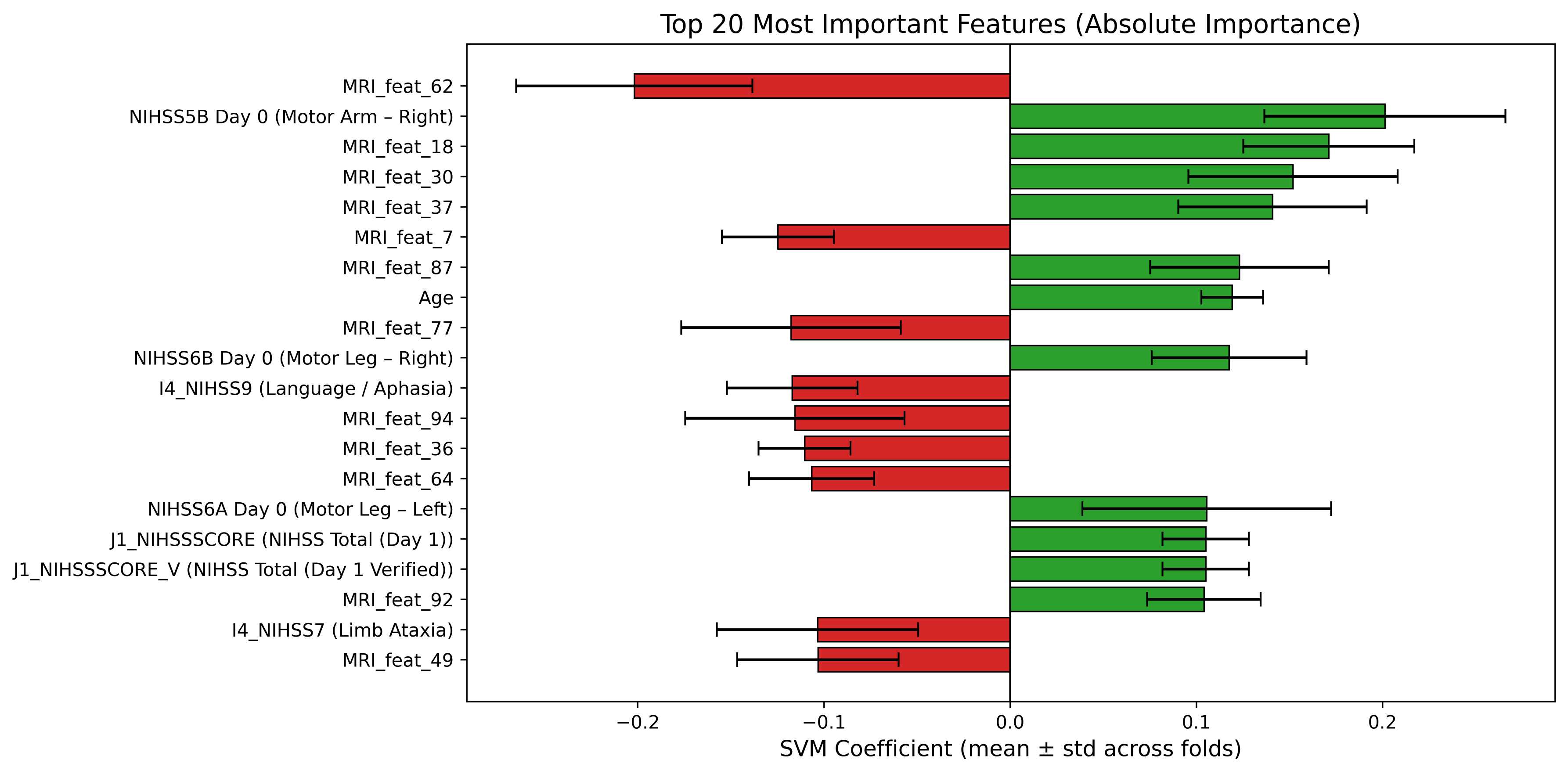}
    \caption{Feature importance derived from linear SVM coefficients}
    \label{fig:feature_importance}
\end{figure}

Feature importance derived from linear SVM coefficients, shown in Figure~\ref{fig:feature_importance}, revealed that both clinical and MRI-derived aspects contribute to outcome prediction. The highest positive weights corresponded to motor NIHSS subscores (Day 0 Right arm motor drift, Day 0 Left arm motor drift) and age, while several MRI-derived embeddings (e.g., MRI\_feat\_18, 30, 37) also rank among the best predictors. Negative weights were dominated by MRI features (e.g., MRI\_feat\_62, 7, 77) and the NIHSS9 (Language/aphasia) subscore. These results indicate that clinical measures of motor impairment and selected diffusion-based latent features jointly drive discrimination between favorable and unfavorable outcomes.

\section{Discussion}
This study shows that integrating quantitative diffusion MRI with structured clinical data enables accurate prediction of three-month outcomes after AIS. Comparison of J0 and J1 ADC imaging demonstrates that J1 models outperform baseline models, confirming that early post-treatment diffusion changes provide key prognostic information. Multimodal fusion with PCA improve model stability.
The better J1 performance reflects the prognostic value of early tissue dynamics related to reperfusion, collateral flow, and secondary injury. The J1 MRI-only model achieves AUC $=0.714\pm0.105$ versus $0.540\pm0.263$ for J0, while the multimodal J1 + clinical + lesion-volume model reaches $0.923\pm0.085$, the highest accuracy consistent with previous findings on 24-h DWI/ADC imaging~\cite{campbell2021acute, kim2023deep}. Clinical variables such as age, pre-stroke mRS, and NIHSS contribute strongly (AUC $=0.882\pm0.105$), confirming their complementary role. Adding lesion-volume improves interpretability and raises AUC by 0.03, highlighting infarct size as an independent predictor. 
PCA stabilizes multimodal fusion by reducing 3D ResNet embeddings to 8–12 orthogonal components ($>$95\% variance), ensuring consistent SVM convergence and offering a reproducible alternative to nonlinear embeddings ~\cite{paszke2019pytorch,pedregosa2011scikit_learn}.

\begin{figure}[t]
    \centering
    \includegraphics[width=1.00\linewidth]{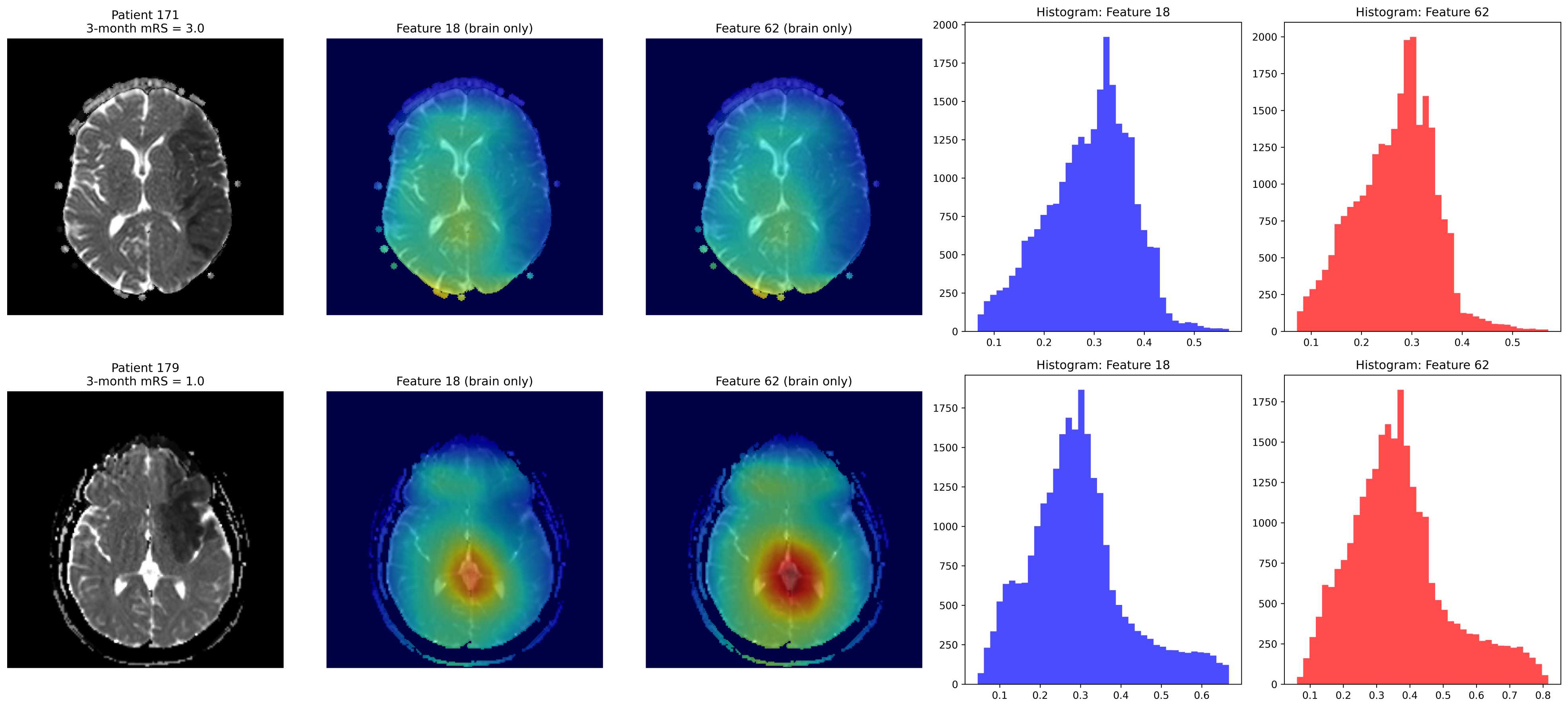}
    \caption{MRI feature maps and histograms for two patients with different three-month mRS scores}
    \label{fig:feature_maps}
\end{figure}

Among clinical predictors, admission NIHSS motor subscores (Day 0 left arm motor drift and right arm motor drift) and age had the highest positive weights, aligning with established prognostic factors. MRI-derived embeddings also contribute but do not yield stable anatomical activation patterns, likely due to the small dataset. Nevertheless, Figure~\ref{fig:feature_maps} demonstrates MRI feature maps for two patients (Patient 171 with mRS = 3.0 and Patient 179 with mRS = 1.0). The gradient-based heat maps show coherent activation within the infarcted hemisphere in severe strokes, suggesting that larger or more homogeneous lesions are better captured. This highlights the complementary roles of structured clinical variables and learned MRI representations in interpretable multimodal prediction.

Unlike end-to-end networks, our PCA–SVM pipeline offers more transparency. Each principal component and SVM coefficient can be inspected, enabling clinicians to trace predictions to specific clinical or image contributors. This makes the framework easier to audit, calibrate and integrate into clinical decision support~\cite{rudin2019stop}.
Early post-treatment MRI enables personalized prognosis by capturing reperfusion success and early tissue evolution, identifying patients at risk of poor recovery despite recanalization. The PCA–SVM pipeline, combining lesion metrics with structured clinical/imaging features, supports transparent translation.
Study limitations include the small sample, automated lesion segmentation and scanner variability. We plan to test the framework in larger cohorts and explore better segmentation, radiomics and multimodality (e.g., perfusion MRI, CT angiography).
\section{Conclusion}
J1 diffusion MRI provides better prognostic value compared with baseline imaging. Integrating deep MRI embeddings with clinical and lesion-volume features (AUC $>$ 0.92) yields a robust, interpretable, and data-efficient model. PCA stabilization enhances reliability and transparency, enabling reproducible outcome prediction. Overall, the proposed PCA–SVM framework offers a clinically translatable foundation for personalized stroke prognosis and will be validated on multicenter longitudinal data.
\section*{Compliance with Ethical Standards}
This retrospective study used anonymized clinical data acquired as part of routine care. According to the institutional guidelines of the University of Geneva and the University Hospital of Geneva, the study is covered under existing retrospective data-use approvals, and no additional protocol-specific ethics approval was required.
The study was conducted in accordance with the principles of the Declaration of Helsinki.
\section*{Reproducibility and Acknowledgments}
This work was supported by the FORDEMS Foundation. 
The authors declare no competing interests. Data and implementation details are available upon request.
\bibliographystyle{IEEEbib}
\bibliography{references}

@article{feigin2021global,
  author =       "V. L. Feigin \textit{et al.}",
  title =        "Global, regional, and national burden of stroke and its risk factors, 1990–2019: a systematic analysis for the Global Burden of Disease Study 2019",
  journal =      "Lancet Neurol.",
  volume =       "20",
  number =       "10",
  pages =        "795--820",
  year =         "2021",
  publisher =    "Elsevier"
}

@article{heiss2017neuroimaging,
  author =       "W. D. Heiss",
  title =        "Contribution of neuro-imaging for prediction of functional recovery after ischemic stroke",
  journal =      "Cerebrovascular Diseases",
  volume =       "44",
  number =       "5-6",
  pages =        "266--276",
  year =         "2017",
  publisher =    "Karger"
}

@article{banks2007stroke,
  author =       "J. L. Banks and C. A. Marotta",
  title =        "Outcomes validity and reliability of the modified Rankin scale: implications for stroke clinical trials: a literature review and synthesis",
  journal =      "Stroke",
  volume =       "38",
  number =       "3",
  pages =        "1091--1096",
  year =         "2007",
  publisher =    "Lippincott Williams & Wilkins"
}

@article{warach1992fast,
  author =       "S. Warach and D. Chien and W. Li and M. Ronthal and R. R. Edelman",
  title =        "Fast magnetic resonance diffusion‐weighted imaging of acute human stroke",
  journal =      "Neurology",
  volume =       "42",
  number =       "9",
  pages =        "1717--1717",
  year =         "1992",
  publisher =    "Lippincott Williams & Wilkins"
}

@article{campbell2021acute,
  author =       "B. C. Campbell \textit{et al.}",
  title =        "Acute stroke imaging research roadmap IV: imaging selection and outcomes in acute stroke clinical trials and practice",
  journal =      "Stroke",
  volume =       "52",
  number =       "8",
  pages =        "2723--2733",
  year =         "2021",
  publisher =    "Lippincott Williams \& Wilkins"
}

@article{goyal2020challenging,
  author =       "M. Goyal \textit{et al.}",
  title =        "Challenging the Ischemic Core Concept in Acute Ischemic Stroke Imaging",
  journal =      "Stroke",
  volume =       "51",
  number =       "10",
  pages =        "3147--3155",
  year =         "2020",
  publisher =    "Lippincott Williams \& Wilkins"
}

@article{fiebach2002ct,
  author =       "J. B. Fiebach \textit{et al.}",
  title =        "CT and diffusion-weighted MR imaging in randomized order: diffusion-weighted imaging results in higher accuracy and lower interrater variability in the diagnosis of hyperacute ischemic stroke",
  journal =      "Stroke",
  volume =       "33",
  number =       "9",
  pages =        "2206--2210",
  year =         "2002",
  publisher =    "Lippincott Williams \& Wilkins"
}

@article{wouters2018prediction,
  author =       "A. Wouters and C. Nysten and V. Thijs and R. Lemmens",
  title =        "Prediction of outcome in patients with acute ischemic stroke based on initial severity and improvement in the first 24 h",
  journal =      "Frontiers in Neurology",
  volume =       "9",
  pages =        "308",
  year =         "2018",
  publisher =    "Frontiers Media"
}

@article{esteva2019guide,
  author =       "A. Esteva \textit{et al.}",
  title =        "A guide to deep learning in healthcare",
  journal =      "Nature Medicine",
  volume =       "25",
  number =       "1",
  pages =        "24--29",
  year =         "2019",
  publisher =    "Nature Publishing Group"
}

@article{nielsen2018prediction,
  author =       "A. Nielsen and M. B. Hansen and A. Tietze and K. Mouridsen",
  title =        "Prediction of tissue outcome and assessment of treatment effect in acute ischemic stroke using deep learning",
  journal =      "Stroke",
  volume =       "49",
  number =       "6",
  pages =        "1394--1401",
  year =         "2018",
  publisher =    "Lippincott Williams \& Wilkins"
}

@article{winzeck2018isles,
  author =       "S. Winzeck \textit{et al.}",
  title =        "ISLES 2016 and 2017-benchmarking ischemic stroke lesion outcome prediction based on multispectral MRI",
  journal =      "Frontiers in Neurology",
  volume =       "9",
  pages =        "679",
  year =         "2018",
  publisher =    "Frontiers Media"
}

@article{liu2023functional,
  author =       "Y. Liu \textit{et al.}",
  title =        "Functional outcome prediction in acute ischemic stroke using a fused imaging and clinical deep learning model",
  journal =      "Stroke",
  volume =       "54",
  number =       "9",
  pages =        "2316--2327",
  year =         "2023",
  publisher =    "Lippincott Williams \& Wilkins"
}

@article{kim2023deep,
  author =       "D. Y. Kim \textit{et al.}",
  title =        "Deep learning-based personalised outcome prediction after acute ischaemic stroke",
  journal =      "Journal of Neurology, Neurosurgery \& Psychiatry",
  volume =       "94",
  number =       "5",
  pages =        "369--378",
  year =         "2023",
  publisher =    "BMJ Publishing Group"
}

@article{lecun2015deep,
  author =       "Y. LeCun and Y. Bengio and G. Hinton",
  title =        "Deep learning",
  journal =      "Nature",
  volume =       "521",
  number =       "7553",
  pages =        "436--444",
  year =         "2015",
  publisher =    "Nature Publishing Group"
}

@inproceedings{ma2024transformer,
  author =       "D. Ma and M. Wang and A. Xiang and Z. Qi and Q. Yang",
  title =        "Transformer-based classification outcome prediction for multimodal stroke treatment",
  booktitle =    "2024 IEEE 2nd International Conference on Sensors, Electronics and Computer Engineering (ICSECE)",
  pages =        "383--386",
  year =         "2024",
  publisher =    "IEEE"
}

@article{wold1987principal,
  author =       "S. Wold and K. Esbensen and P. Geladi",
  title =        "Principal component analysis",
  journal =      "Chemometrics and Intelligent Laboratory Systems",
  volume =       "2",
  number =       "1-3",
  pages =        "37--52",
  year =         "1987",
  publisher =    "Elsevier"
}

@article{cardoso2022monai,
  author =       "M. J. Cardoso \textit{et al.}",
  title =        "Monai: An open-source framework for deep learning in healthcare",
  journal =      "arXiv preprint arXiv:2211.02701",
  year =         "2022",
  url =          "https://arxiv.org/abs/2211.02701"
}

@article{chen2019med3d,
  author =       "S. Chen and K. Ma and Y. Zheng",
  title =        "Med3D: Transfer learning for 3D medical image analysis",
  journal =      "arXiv preprint arXiv:1904.00625",
  year =         "2019",
  url =          "https://arxiv.org/abs/1904.00625"
}

@article{wilcoxon1945individual,
  author =       "F. Wilcoxon",
  title =        "Individual comparisons by ranking methods",
  journal =      "Biometrics Bulletin",
  volume =       "1",
  number =       "6",
  pages =        "80--83",
  year =         "1945",
  publisher =    "International Biometric Society"
}

@article{cortes1995support,
  author =       "C. Cortes and V. Vapnik",
  title =        "Support-vector networks",
  journal =      "Machine Learning",
  volume =       "20",
  number =       "3",
  pages =        "273--297",
  year =         "1995",
  publisher =    "Springer"
}

@article{pedregosa2011scikit_learn,
  author =       "F. Pedregosa \textit{et al.}",
  title =        "Scikit-learn: Machine learning in Python",
  journal =      "The Journal of Machine Learning Research",
  volume =       "12",
  pages =        "2825--2830",
  year =         "2011",
  publisher =    "JMLR.org"
}

@article{paszke2019pytorch,
  author =       "A. Paszke \textit{et al.}",
  title =        "PyTorch: An imperative style, high-performance deep learning library",
  journal =      "Advances in Neural Information Processing Systems",
  volume =       "32",
  year =         "2019"
}

@article{rudin2019stop,
  author =       "C. Rudin",
  title =        "Stop explaining black box machine learning models for high stakes decisions and use interpretable models instead",
  journal =      "Nature Machine Intelligence",
  volume =       "1",
  number =       "5",
  pages =        "206--215",
  year =         "2019",
  publisher =    "Nature Publishing Group"
}
\end{document}